\documentclass[11pt,preprintnumbers,titlepage,nofootinbib]{revtex4-2}
\usepackage[latin9]{inputenc}
\usepackage{amsmath}
\usepackage{amssymb}
\usepackage{graphicx}
\usepackage[unicode=true,
 bookmarks=false,
 breaklinks=false,pdfborder={0 0 1},backref=section,colorlinks=false]
 {hyperref}
\hypersetup{
 colorlinks,linkcolor=blue,citecolor=blue,urlcolor=blue}

\makeatletter

\usepackage{wasysym}
\usepackage{amsfonts}
\usepackage{subfigure}
\usepackage{cleveref}
\usepackage{listings}
\usepackage{xcolor}
\usepackage{array}
\usepackage{url}
\usepackage{color}

\@ifundefined{textcolor}{}{
\definecolor{BLACK}{gray}{0}
\definecolor{WHITE}{gray}{1}
\definecolor{RED}{rgb}{1,0,0}
\definecolor{GREEN}{rgb}{0,1,0}
\definecolor{BLUE}{rgb}{0,0,1}
\definecolor{CYAN}{cmyk}{1,0,0,0}
\definecolor{MAGENTA}{cmyk}{0,1,0,0}
\definecolor{YELLOW}{cmyk}{0,0,1,0}
}

\makeatother

\begin{document}
\preprint{CTP-SCU/2022009}
\title{Appearance of an Infalling Star in Black Holes with Multiple Photon
Spheres}
\author{Yiqian Chen$^{a}$}
\email{chenyiqian@stu.scu.edu.cn}

\author{Guangzhou Guo$^{a}$}
\email{guangzhouguo@stu.scu.edu.cn}

\author{Peng Wang$^{a}$}
\email{pengw@scu.edu.cn}

\author{Houwen Wu$^{a,b}$}
\email{hw598@damtp.cam.ac.uk}

\author{Haitang Yang$^{a}$}
\email{hyanga@scu.edu.cn}

\affiliation{$^{a}$Center for Theoretical Physics, College of Physics, Sichuan
University, Chengdu, 610064, China}
\affiliation{$^{b}$Department of Applied Mathematics and Theoretical Physics,
University of Cambridge, Wilberforce Road, Cambridge, CB3 0WA, UK}
\begin{abstract}
Photon spheres play a pivotal role in the imaging of luminous objects
near black holes. In this paper, we examine observational appearances
of a star freely falling in hairy black holes, which can possess one
or two photon spheres outside the event horizon. When there exists
a single photon sphere, the total luminosity measured by distant observers
decreases exponentially with time at late times. Due to successive
arrivals of photons orbiting around the photon sphere different times,
a specific observer would see a series of light flashes with decreasing
intensity, which share a similar frequency content. Whereas in the
case with two photon spheres, photons temporarily trapped between
the photon spheres can cause a peak of the total luminosity, which
is followed by a slow exponential decay, at late times. In addition,
these photons lead to one more series of light flashes seen by the
specific observer.\ 
\end{abstract}
\maketitle
\tableofcontents{}

\section{Introduction}

\label{sec:Introduction}

The recent announcements of the images of the supermassive black holes
M87{{*} \cite{Akiyama:2019cqa,Akiyama:2019brx,Akiyama:2019sww,Akiyama:2019bqs,Akiyama:2019fyp,Akiyama:2019eap,Akiyama:2021qum,Akiyama:2021tfw}
and }Sgr A{*} \cite{EventHorizonTelescope:2022xnr,EventHorizonTelescope:2022vjs,EventHorizonTelescope:2022wok,EventHorizonTelescope:2022exc,EventHorizonTelescope:2022urf,EventHorizonTelescope:2022xqj}
by the Event Horizon Telescope (EHT) collaboration open a new window
to test general relativity in the strong field regime. These images
display two main features, a central brightness depression, dubbed
``shadow'', and a bright ring, which is closely relevant to a class
of circular and unstable photon orbits. For static spherically symmetric
black holes, these circular and unstable photon orbits would form
photon spheres outside the event horizon. Since the shadow and the
bright ring are originated from light deflections by the strong gravitational
field near the photon spheres \cite{Synge:1966okc,Bardeen:1972fi,Bardeen:1973tla,Virbhadra:1999nm,Claudel:2000yi,Virbhadra:2008ws,Bozza:2009yw,Virbhadra:2022iiy},
black hole images can encode valuable information of the geometry
in the vicinity of the photon spheres. In particular, black hole images
have been widely studied in the context of different theories of gravity,
e.g., nonlinear electrodynamics \cite{Atamurotov:2015xfa,Stuchlik:2019uvf,Ma:2020dhv,Hu:2020usx,Kruglov:2020tes,Zhong:2021mty,He:2022opa},
the Gauss-Bonnet theory \cite{Ma:2019ybz,Wei:2020ght,Zeng:2020dco,Guo:2020zmf},
fuzzball \cite{Bacchini:2021fig}, the Chern-Simons type theory \cite{Ayzenberg:2018jip,Amarilla:2010zq},
$f(R)$ gravity \cite{Dastan:2016vhb,Addazi:2021pty,Li:2021ypw},
string inspired black holes \cite{Amarilla:2011fx,Guo:2019lur,Zhu:2019ura,Kumar:2020hgm},
wormholes \cite{Shaikh:2018oul,Wielgus:2020uqz,Guerrero:2021ues,Peng:2021osd,Bambi:2021qfo,Olmo:2021piq,Guerrero:2022qkh},
bosonic stars \cite{Cunha:2017wao,Olivares:2018abq} and other theories
\cite{Wei:2013kza,Wang:2017hjl,Mizuno:2018lxz,Zhang:2020xub,Zeng:2021mok,Guo:2021wid,Meng:2022kjs,Qiao:2022jlu,Bogush:2022hop,Guo:2022rql,Wang:2022yvi,Zhang:2022klr,Hou:2022gge}.
Furthermore, the EHT observations can also be applied to impose constraints
on the cosmological parameters \cite{Perlick:2018iye,Tsupko:2019pzg,Vagnozzi:2020quf,Li:2020drn,Roy:2020dyy,Chowdhuri:2020ipb}
and the size of extra dimensions \cite{Amir:2017slq,Vagnozzi:2019apd,Banerjee:2019nnj},
test the equivalence principle \cite{Li:2019lsm,Gralla:2020pra,Li:2021mzq},
and probe some fundamental physics issues including dark matter \cite{Chen:2019fsq,Konoplya:2019sns,Jusufi2019,Davoudiasl2019,Roy:2019esk,Saurabh:2020zqg}
and dark energy \cite{Abdujabbarov:2015pqp,Zeng:2020vsj,Qin:2020xzu,Yuan-Chen:2020sek,He:2021aeo}.

Apart from stationary observational appearances of black holes, photon
spheres also play an important role in dynamic observations of luminous
objects around black holes. A particularly interesting scenario is
luminous matter freely falling onto black holes, which occurs frequently
and was reported near the Cyg X-1 black hole \cite{Dolan:2011vz}
and the Sgr A{*} source \cite{GRAVITY:2020lpa,GRAVITY:2020xcu}. Recently,
the authors of \cite{Cardoso:2021sip} numerically simulated the appearance
of a point-like source, which emits photons or gravitons as it falls
into a Schwarzschild black hole. Note that there exists a single photon
sphere outside the event horizon in a Schwarzschild black hole. It
showed that, owing to radiations temporarily trapped just outside
the photon sphere, late-time radiations are dominated by blueshifted
photons or gravitons with a near-critical impact parameter, and the
total luminosity decreases exponentially as $e^{-\lambda t}$, where
$\lambda$ is the Lyapunov exponent of circular null geodesics at
the photon sphere.

On the other hand, a novel class of Einstein-Maxwell-scalar (EMS)
models have recently been constructed to understand the formation
of hairy black holes \cite{Herdeiro:2018wub,Konoplya:2019goy,Wang:2020ohb,Guo:2021zed,Guo:2021ere}.
In such models, the scalar field non-minimally couples to the electromagnetic
field, which can trigger a tachyonic instability to form spontaneously
scalarized hairy black holes from Reissner-Nordström (RN) black holes.
Properties of the hairy black holes have been extensively studied
in the literature, e.g., different non-minimal coupling functions
\cite{Fernandes:2019rez,Fernandes:2019kmh,Blazquez-Salcedo:2020nhs},
massive and self-interacting scalar fields \cite{Zou:2019bpt,Fernandes:2020gay},
horizonless reflecting stars \cite{Peng:2019cmm}, stability analysis
\cite{Myung:2018vug,Myung:2019oua,Zou:2020zxq,Myung:2020etf,Mai:2020sac},
higher dimensional scalar-tensor models \cite{Astefanesei:2020qxk},
quasinormal modes \cite{Myung:2018jvi,Blazquez-Salcedo:2020jee},
two U$\left(1\right)$ fields \cite{Myung:2020dqt}, quasi-topological
electromagnetism \cite{Myung:2020ctt}, topology and spacetime structure
influences \cite{Guo:2020zqm}, and in dS/AdS spacetime \cite{Brihaye:2019dck,Brihaye:2019gla,Zhang:2021etr,Guo:2021zed}.

Intriguingly, the scalarized hairy black holes have been shown to
possess two photon spheres outside the event horizon in certain parameter
regions \cite{Gan:2021pwu,Gan:2021xdl}. The existence of an extra
photon sphere can significantly affect optical appearances of surrounding
accretion disks (e.g., producing bright rings of different radius
in the black hole images \cite{Gan:2021pwu} and noticeably increasing
the flux of the observed images \cite{Gan:2021xdl}) and luminous
celestial spheres (e.g., tripling higher-order images of the celestial
spheres \cite{Guo:2022muy}). Moreover, the effective potential for
a scalar perturbation in the hairy black holes with two photon spheres
was found to exhibit a double-peak structure, leading to long-lived
quasinormal modes \cite{Guo:2021enm} and echo signals \cite{Guo:2022umh}.
It is worth noting that the existence of two photon spheres outside
the event horizon has also been reported for dyonic black holes with
a quasi-topological electromagnetic term \cite{Liu:2019rib}, black
holes in massive gravity \cite{deRham:2010kj,Dong:2020odp} and wormholes
in the black-bounce spacetime \cite{Tsukamoto:2021caq,Tsukamoto:2021fpp,Tsukamoto:2022vkt}.

In this paper, we aim to study how an extra photon sphere affects
observational appearances of a freely-falling star in the context
of the hairy black holes. The rest of the paper is organized as follows.
In Section \ref{sec:Set up}, we briefly review hairy black hole solutions
in the EMS model and introduce the observational settings. Numerical
results are presented in Section \ref{sec:numerical Results}. Section
\ref{sec:CONCLUSIONS} is devoted to our main conclusions. We set
$16\pi G=c=1$ throughout the paper.

\section{Set up}

\label{sec:Set up}

We consider a 4-dimensional EMS model with the action \cite{Herdeiro:2018wub}
\begin{equation}
S=\int d^{4}x\sqrt{-g}\left[\mathcal{R}-2\partial_{\mu}\phi\partial^{\mu}\phi-e^{\alpha\phi^{2}}F_{\mu\nu}F^{\mu\nu}\right],\label{eq:Action}
\end{equation}
where $\mathcal{R}$ is the Ricci scalar, and $F_{\mu\nu}=\partial_{\mu}A_{\nu}-\partial_{\nu}A_{\mu}$
is the electromagnetic field strength tensor. In this EMS model, the
scalar field $\phi$ is non-minimally coupled to the electromagnetic
field $A_{\mu}$ with the coupling function $e^{\alpha\phi^{2}}$.
Restricting to static and spherically symmetric black hole solutions,
one can have the generic ansatz 
\begin{align}
ds^{2} & =-N(r)e^{-2\delta(r)}dt^{2}+\frac{1}{N(r)}dr^{2}+r^{2}\left(d\theta^{2}+\sin^{2}\theta d\varphi^{2}\right),\nonumber \\
A_{\mu}dx^{\mu} & =V(r)dt\text{ and}\ \phi=\phi(r).\label{eq:HBH}
\end{align}
For this ansatz, the equations of motion are 
\begin{align}
N^{\prime}(r) & =\frac{1-N(r)}{r}-\frac{Q^{2}}{r^{3}e^{\alpha\phi^{2}(r)}}-rN(r)\left[\phi^{\prime}(r)\right]^{2},\nonumber \\
\left[r^{2}N(r)\phi^{\prime}(r)\right]^{\prime} & =-\frac{\alpha Q^{2}\phi(r)}{r^{2}e^{\alpha\phi^{2}(r)}}-r^{3}N(r)\left[\phi^{\prime}(r)\right]^{3},\nonumber \\
\delta^{\prime}(r) & =-r\left[\phi^{\prime}(r)\right]^{2},\label{eq:EOM}\\
V^{\prime}(r) & =\frac{Q}{r^{2}e^{\alpha\phi^{2}(r)}}e^{-\delta(r)},\nonumber 
\end{align}
where the integration constant $Q$ can be interpreted as the electric
charge of the black hole, and primes stand for derivative with respect
to $r$. In addition, the ADM mass $M$ can be obtained via $M=\lim\limits _{r\rightarrow\infty}r\left[1-N(r)\right]/2$.
Moreover, we impose proper boundary conditions at the event horizon
$r_{h}$ and spatial infinity, 
\begin{align}
N(r_{h}) & =0\text{, }\delta(r_{h})=\delta_{0}\text{, }\phi(r_{h})=\phi_{0}\text{, }V(r_{h})=0\text{,}\nonumber \\
N(\infty) & =1\text{,}\ \delta(\infty)=0\text{, }\phi(\infty)=0\text{, }V(\infty)=\Phi\text{,}\label{eq:boundary conditions}
\end{align}
where $\delta_{0}$ and $\phi_{0}$ can be used to characterize black
hole solutions, and $\Phi$ is the electrostatic potential. Specifically,
$\phi_{0}=\delta_{0}=0$ correspond to the scalar-free solutions with
$\phi=0$, i.e., RN black holes. When non-zero values of $\phi_{0}$
and $\delta_{0}$ are admitted, hairy black hole solutions with a
non-trivial scalar field $\phi$ are obtained by a shooting method
built in the \textit{$NDSolve$} function of \textit{$Wolfram\text{ }\circledR Mathematica$}.

In this paper, we study a point-like star freely falling along the
radial direction at $\theta=\frac{\pi}{2}$ and $\varphi=0$. The
equations governing geodesics on the equatorial plane can be derived
from the Lagrangian 
\begin{equation}
\mathcal{L}=\frac{1}{2}\left[-N(r)e^{-2\delta(r)}\dot{t}^{2}+\frac{1}{N(r)}\dot{r}^{2}+r^{2}\dot{\varphi}^{2}\right],\label{eq:Geo Lagrangian}
\end{equation}
where dots stand for derivative with respect to some affine parameter
$\lambda$. Note that $\mathcal{L}=-1/2$ for time-like geodesics
if $\lambda$ is the proper time. Since $t$ and $\varphi$ do not
explicitly appear in the Lagrangian, one can characterize geodesics
by their energy $E$ and angular momentum $L$, namely 
\begin{equation}
p_{t}\equiv-N(r)e^{-2\delta(r)}\dot{t}=-E\text{, }\quad p_{\varphi}\equiv r^{2}\dot{\varphi}=L.\label{eq:E-L}
\end{equation}
For simplicity, we specialize to a radially freely falling star starting
at infinity with a zero initial velocity, whose four-velocity is 
\begin{equation}
v_{e}^{\mu}(r)=\left(\frac{e^{2\delta(r)}}{N(r)},-\sqrt{e^{2\delta(r)}-N(r)},0,0\right).\label{eq:ve}
\end{equation}

Suppose that the star emits photons when it falls radially onto a
black hole. Due to spherical symmetry, we can confine ourselves to
emissions on the equatorial plane. The Lagrangian $\left(\ref{eq:Geo Lagrangian}\right)$
can also be used to describe null geodesics on the equatorial plane.
Moreover, the constancy of the Lagrangian $\mathcal{L}=0$ rewrites
the radial component of the null geodesic equations as 
\begin{equation}
\frac{e^{-2\delta(r)}}{L^{2}}\dot{r}^{2}=\frac{1}{b^{2}}-V_{\text{eff}}(r),\label{eq:Pr}
\end{equation}
where $b\equiv L/E$ is the impact parameter, and $V_{\text{eff}}\equiv e^{-2\delta(r)}N(r)r^{-2}$
is the effective potential. Unstable circular null geodesics at radius
$r_{\text{ph}}$, which constitute a photon sphere of radius $r_{\text{ph}}$,
are determined by 
\begin{equation}
V_{\text{eff}}\left(r_{\text{ph}}\right)=\frac{1}{b_{\text{ph}}^{2}},\quad V_{\text{eff}}^{\prime}\left(r\right)=0,\quad V_{\text{eff}}^{\prime\prime}\left(r_{\text{ph}}\right)<0,
\end{equation}
where $b_{\text{ph}}$ is the corresponding impact parameter. In short,
a local maximum of the effective potential corresponds to a photon
sphere. Remarkably, it showed that, when the black hole charge is
large enough, $V_{\text{eff}}(r)$ of the hairy black hole solutions
$\left(\ref{eq:HBH}\right)$ can have two local maxima, leading to
a double-peak structure \cite{Gan:2021pwu,Gan:2021xdl}. Consequently,
the hairy black holes with the double-peak effective potential possess
two photon spheres outside the event horizon. Indeed, the effective
potential of the hairy black hole with $\alpha=0.9$ and $Q=1.064M$
is plotted in FIG. \ref{Fig:Potential-d}, which exhibits a double-peak
structure.

To trace light rays from the star to a far-away observers, one need
to supply initial conditions for Eqs. $\left(\ref{eq:E-L}\right)$
and $\left(\ref{eq:Pr}\right)$. For a photon of four-momentum $p_{\mu}$,
the momentum measured in the rest frame of the star at $r=r_{e}$
is 
\begin{align}
p^{\hat{t}} & =-\left[N(r_{e})e^{-2\delta(r_{e})}\right]^{-1}p_{t}+\sqrt{e^{2\delta(r_{e})}-N(r_{e})}p_{r},\nonumber \\
p^{\hat{r}} & =-\left[N(r_{e})e^{-2\delta(r_{e})}\right]^{-1}\sqrt{1-N(r_{e})e^{-2\delta(r_{e})}}p_{t}+\sqrt{e^{2\delta(r_{e})}}p_{r},\label{eq:pR}\\
p^{\hat{\theta}} & =0\text{, }p^{\hat{\varphi}}=\frac{p_{\varphi}}{r_{e}}.\nonumber 
\end{align}
The emission angle $\beta$ is defined as 
\begin{equation}
\cos\beta=\frac{p^{\hat{r}}}{p^{\hat{t}}},\label{eq:alpha}
\end{equation}
which is the angle between the propagation direction of the photon
and the radial direction in the rest frame of the star. With Eqs.
$\left(\ref{eq:pR}\right)$ and $\left(\ref{eq:alpha}\right)$, one
has the normalized frequency of the photon measured by the far-away
observer, 
\begin{equation}
\frac{\omega_{o}}{\omega_{e}}\equiv\frac{p^{t}}{p^{\hat{t}}}=1-\cos\beta\sqrt{1-N(r_{e})e^{-2\delta(r_{e})}},\label{eq:wratio}
\end{equation}
where $\omega_{e}$ and $\omega_{o}$ are rest-frame and observed
photon frequencies, respectively. Furthermore, the luminosity is defined
as $L_{i}\equiv dE_{i}/d\tau_{i}$, where $E_{i}$ is the total energy
of the photon, $\tau_{i}$ is the proper time, and $i=e$ and $o$
denote quantities corresponding to the emitter and observer, respectively.
Similar to the normalized frequency, we define the normalized luminosity
as 
\[
\frac{L_{o}}{L_{e}}=\frac{dE_{o}/d\tau_{o}}{dE_{e}/d\tau_{e}}\approx\frac{\omega_{o}dn_{o}}{\omega_{e}dn_{e}}\left(\frac{dt_{o}}{d\tau_{e}}\right)^{-1},
\]
where $n_{o}$ and $n_{e}$ are the observed and emitted photon numbers,
respectively, and we replaced $d\tau_{o}$ by $dt_{o}$ since they
are almost the same for the far-away observer.

\section{Numerical Results}

\label{sec:numerical Results}

In this section, we numerically study observational appearances of
a star falling radially towards hairy black holes with a single or
two photon spheres outside the event horizon. During the free fall
of the star, photons are emitted isotropically in its rest frame.
Specifically, we assume that the star starts emitting photons at $t=0$
and $r_{e}=31.155M$, and emits $3200$ photons, which are uniformly
distributed in the emission angle $\beta$, every proper time interval
$\delta\tau=0.002M$. In addition, observers collect photons on a
celestial sphere at the radius $r_{o}=100M$. To calculate observed
luminosities, the collected photons are grouped into packets of 50
(i.e., $dn_{o}=50$) according to their arrival time. Without loss
of generality, we set $M=1$ in the rest of this section.

\subsection{Single-peak Potential}

\label{sub-sec:single peak}

\begin{figure}[ptb]
\includegraphics[width=0.6\textwidth]{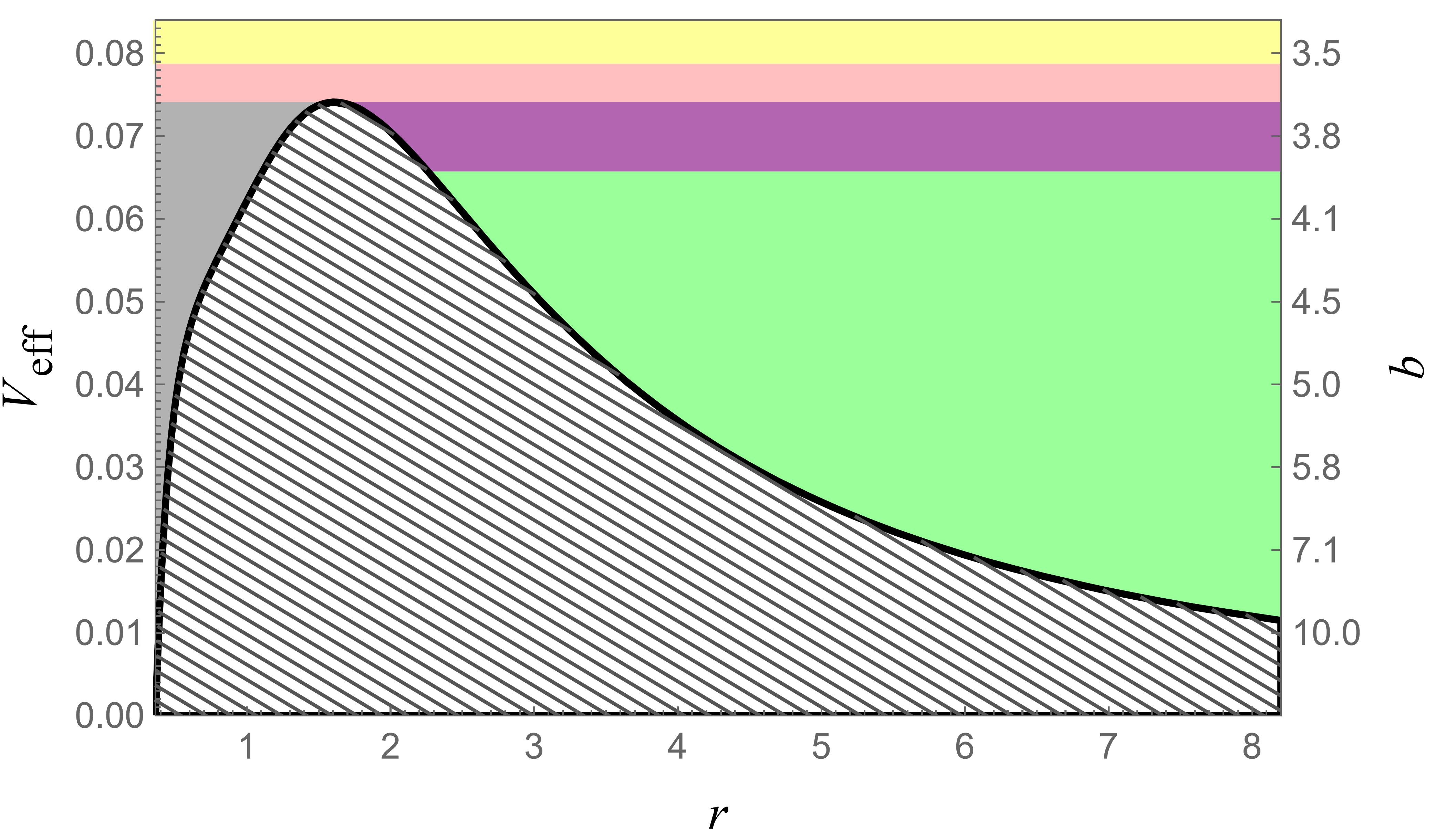}\caption{The black line denotes the effective potential of null geodesics in
the hairy black hole with $\alpha=0.9$ and $Q=1.054$. There is only
one potential maximum at $r_{\text{ph}}=1.610$, corresponding to
the photon sphere with the critical impact parameter $b_{\text{ph}}=3.673$.
Colored regions are presented in the $r$-$b$ parameter space, for
which a photon with the impact parameter $b$ is emitted at $r$.
A far-away observer can not receive photons emitted in the gray region.
Photons emitted in the pink and purple regions have impact parameters
close to $b_{\text{ph}}$, hence can be temporarily trapped around
the photon sphere. In particular, after photons in the pink (purple)
region are emitted outward inside (inward outside) the photon sphere,
they would circle around the photon sphere more than once before reaching
an observer.}
\label{Fig:Potential-s}
\end{figure}

\begin{figure}[ptb]
\includegraphics[width=1\textwidth]{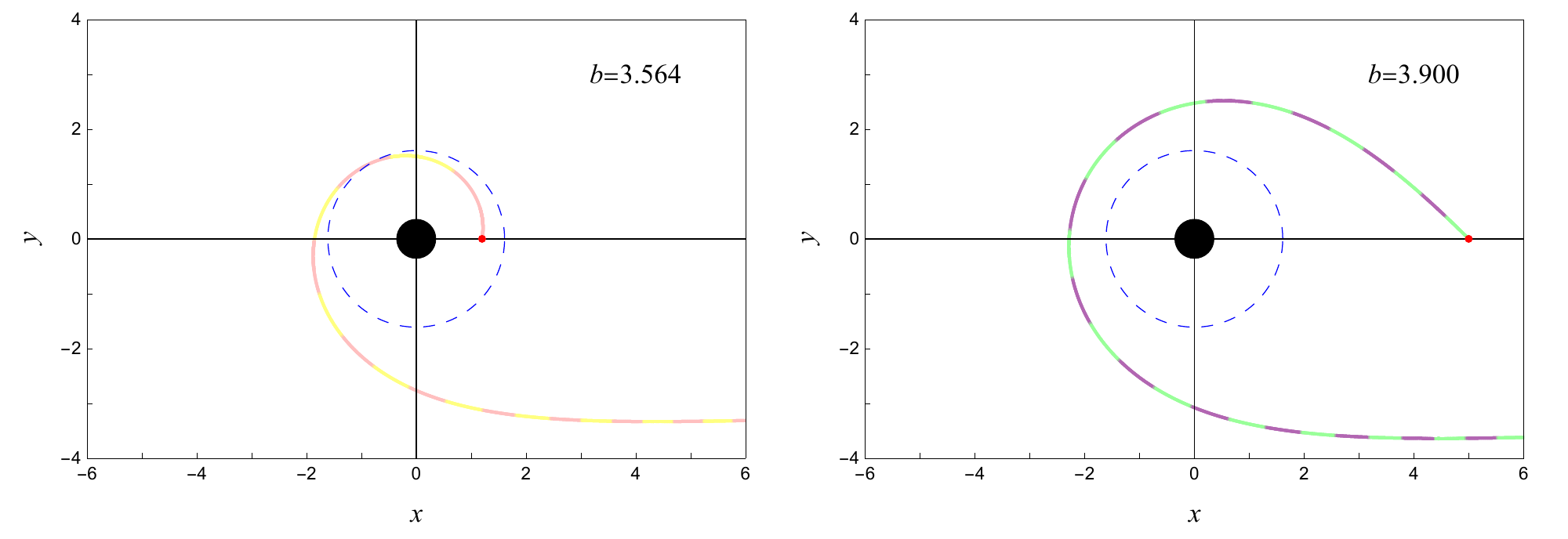}\caption{Photon trajectories in the hairy black hole with a single potential.
The blue dashed circles denote the photon sphere. \textbf{Left}: The
trajectory of a photon with $b=3.564$ emitted at $r_{e}=1.2$. The
light ray has the angular coordinate change $\Delta\varphi=2\pi$
when reaching a celestial sphere at $r_{o}=100$. \textbf{Right}:
The trajectory of a photon with $b=3.900$ emitted at $r_{e}=5$.
The light ray has the angular coordinate change $\Delta\varphi=2\pi$
when reaching the celestial sphere.}
\label{Fig:tr-s}
\end{figure}

\begin{figure}[ptb]
\includegraphics[width=1\textwidth]{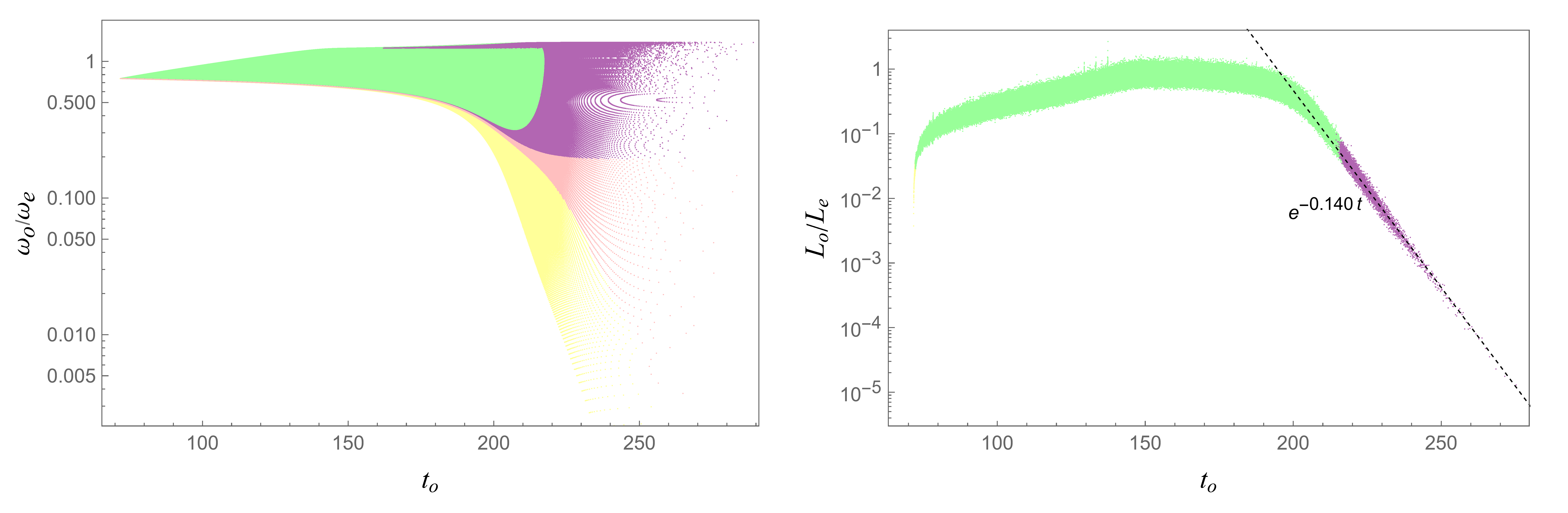}\caption{The normalized frequency distribution and total luminosity of photons
emitted from a freely-falling star in the hairy black hole with a
single-peak potential. The star radially falls along $\phi=0$ on
the the equatorial plane from spatial infinity at rest and starts
emitting isotropically in its rest frame at $r_{e}=31.155$. The emitted
photons are collected by observers on a celestial sphere at $r_{o}=100$.
\textbf{Left:} The observers receive photons with a wide range of
frequencies. In the early stage, photons emitted in the green region
of FIG. \ref{Fig:Potential-s} dominate the frequency observation.
The late-time high-frequency observation is determined by photons
emitted in the purple region, which undergo extreme bending near the
photon sphere. \textbf{Right:} The luminosity is calculated by grouping
received photons into packets of 50. At late times, the luminosity
is mainly controlled by photons emitted with a near-critical impact
parameter in the purple region. The luminosity decays exponentially
as $e^{-\lambda t}\simeq e^{-0.140t}$, where $\lambda$ is the Lyapunov
exponent of circular null geodesics at the photon sphere. Note that
the imaginary part of the associated lowest-lying quasinormal modes
is $-\lambda/2$.}
\label{Fig:sky-s}
\end{figure}

\begin{figure}[ptb]
\includegraphics[width=1\textwidth]{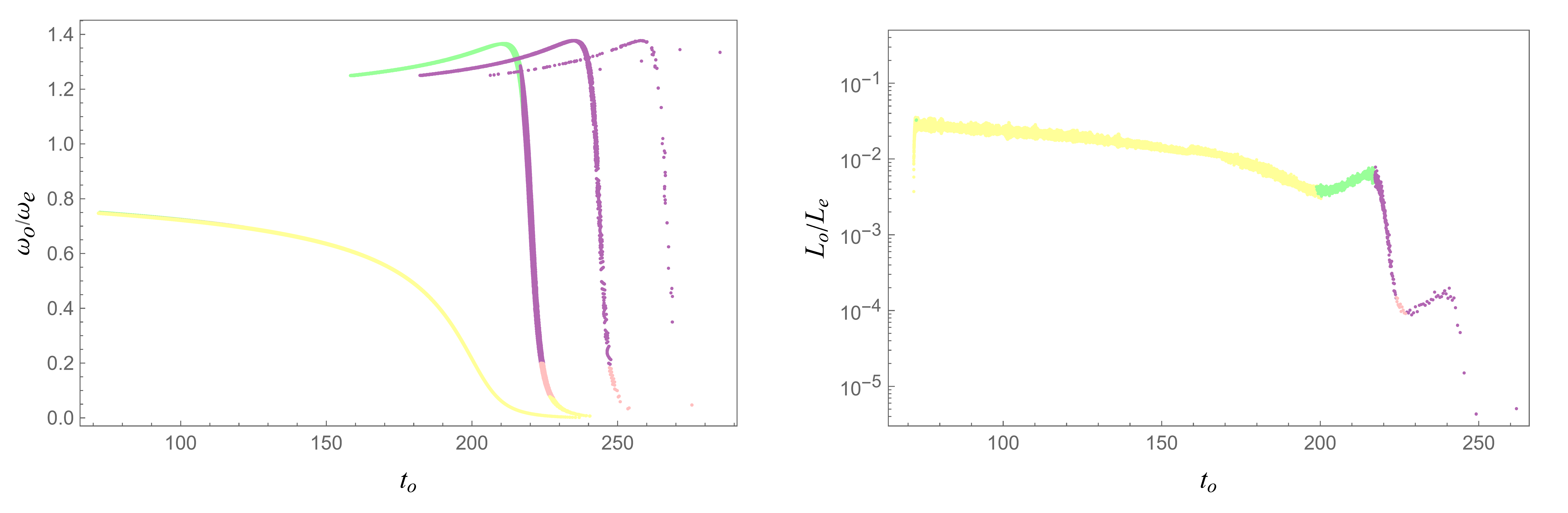}\caption{The normalized frequency and luminosity of the infalling star measured
by a far-away observer at $r_{o}=100$, $\theta=\pi/2$ and $\phi=0$
in the hairy black hole with a single-peak potential. The colored
dots denote photons emitted in the regions with the same color in
FIG. \ref{Fig:Potential-s}. \textbf{Left:} The received photons form
several frequency lines indexed by the number $n$ of photon orbits
around the black hole. From left to right, the frequency lines correspond
to $n=0$, $1$, $2$ and $3$, respectively. In particular, photons
with a impact parameter close to $b_{\text{ph}}$ give rise to the
self-similar $n\protect\geq1$ lines, which display blueshifts caused
by the Doppler effect and the strong gravitational lensing. The time
delay between the $n\protect\geq1$ lines is roughly the period of
circular null geodesics at the photon sphere, $\Delta T\simeq2\pi b_{\text{ph}}\simeq23$.
\textbf{Right: }At early times, the luminosity is dominated by outward-emitted
photons with a small impact parameter and gradually decreases. Subsequently,
blueshifted $n=1$ photons start to reach the observer and then become
the most dominant contribution, which results in the growth of the
luminosity. Later, the luminosity drops sharply with the decrease
in the frequency of the $n=1$ photons. Consequently, a flash is observed
around $t_{o}\simeq215$. When the $n=2$ photons play out, a similar
steady rise followed by a steep fall is also observed, leading to
a much fainter flash.}
\label{Fig:phio-s}
\end{figure}

We first consider the hairy black hole with $\alpha=0.9$ and $Q=1.054$,
which has a single-peak potential as shown in FIG. \ref{Fig:Potential-s}.
The photon sphere is located at $r_{\text{ph}}=1.610$, and the corresponding
critical impact parameter is $b_{\text{ph}}=3.673$. Ingoing photons
with $b$ $<b_{\text{ph}}$ can overcome the effective potential barrier
and get captured by the black hole. Therefore for $b$ $<b_{\text{ph}}$,
a distant observer can only receive outward-emitted photons. On the
other hand, inward-emitted photons with $b$ $>b_{\text{ph}}$ can
be observed since they bounce outward from the potential barrier.
Interestingly, strong gravitational lensing near the photon sphere
causes extreme bending of light rays with a near-critical impact parameter,
which makes them temporarily trapped around the photon sphere and
take a long time to reach the observer. As a result, light rays with
a near-critical impact parameter play a crucial role in the observational
appearance of the infalling star at late times. To illustrate photons
with a near-critical impact parameter, we classify photons into four
categories according to their $b$,
\begin{itemize}
\item $b<3.564$. Yellow region in FIG. \ref{Fig:Potential-s} and yellow
dots in FIGs. \ref{Fig:sky-s} and \ref{Fig:phio-s}.
\item $3.564\leq b<b_{\text{ph}}$. Pink region in FIG. \ref{Fig:Potential-s}
and pink dots in FIGs. \ref{Fig:sky-s} and \ref{Fig:phio-s}. The
left panel of FIG. \ref{Fig:tr-s} depicts a light ray of $b=3.564$,
which corresponds to an outward-emitted photon that is emitted at
$r_{e}=1.2$ inside the photon sphere and reaches the celestial sphere
after the change of angular coordinate $\Delta\varphi=2\pi$. It is
worth emphasizing that near-critical photons of $b<b_{\text{ph}}$
that are emitted outward inside the photon sphere would linger near
the photon sphere for some time.
\item $b_{\text{ph}}<b\leq3.900$. Purple region in FIG. \ref{Fig:Potential-s}
and purple dots in FIGs. \ref{Fig:sky-s} and \ref{Fig:phio-s}. The
right panel of FIG. \ref{Fig:tr-s} shows the trajectory of an inward-emitted
photon with $b=3.900$, which is emitted at $r_{e}=5$ outside the
photon sphere and reaches the celestial sphere after the change of
angular coordinate $\Delta\varphi=2\pi$. Near-critical photons of
$b>b_{\text{ph}}$ that are emitted inward can circle around the photon
sphere more than once before being observed, and contribute significantly
to late-time observations. In this case, the photon sphere acts as
a reflecting wall to scatter photons, which indicates that the near-critical
photons can undergo a blueshift.
\item $b>3.900$. Green region in FIG. \ref{Fig:Potential-s} and green
dots in FIGs. \ref{Fig:sky-s} and \ref{Fig:phio-s}. 
\end{itemize}
\begin{figure}[ptb]
\includegraphics[width=0.5\textwidth]{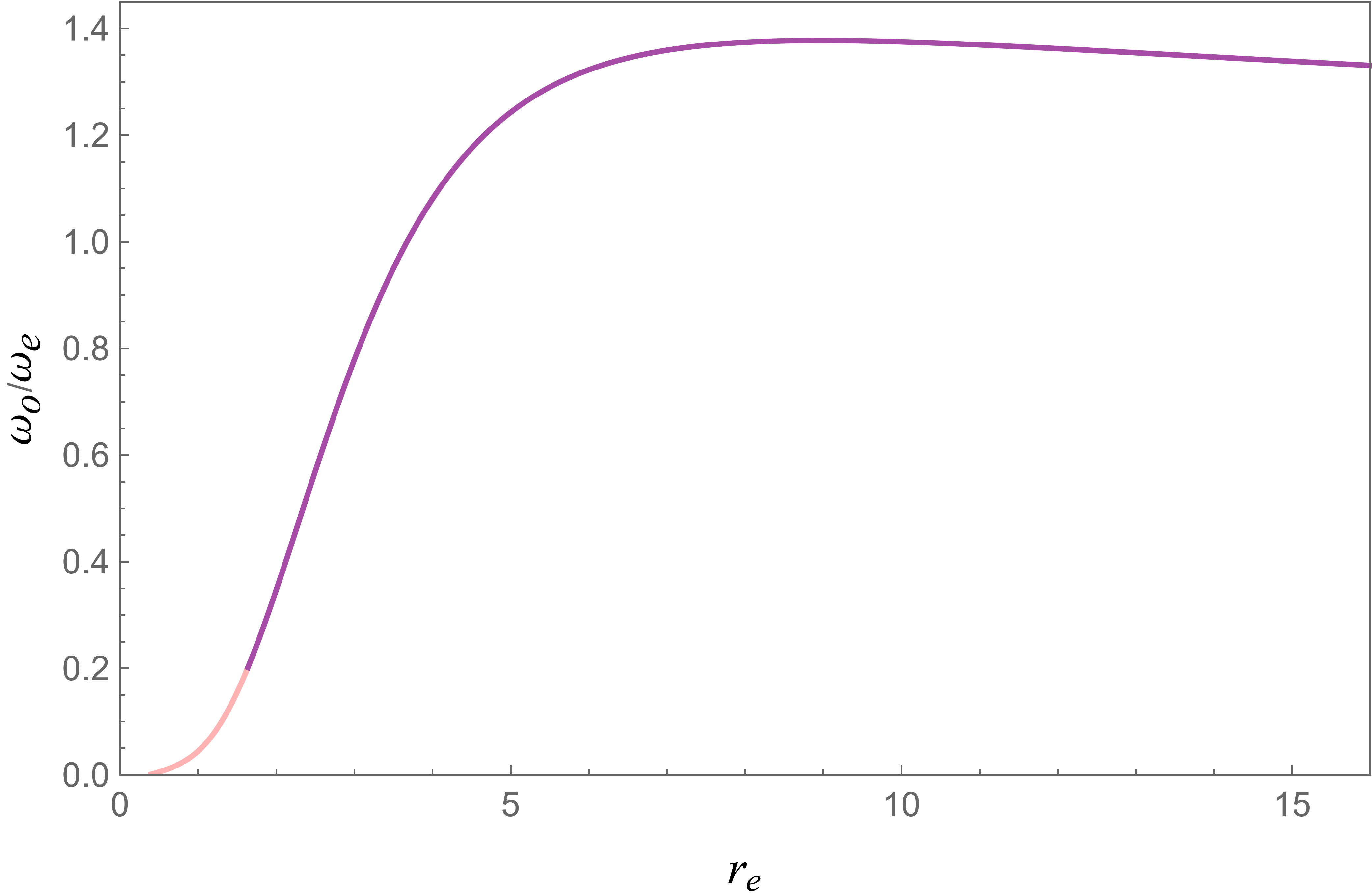}\caption{The normalized frequency $\omega_{o}/\omega_{e}$ as a function of
the position of the star $r_{e}$ for photons of impact parameter
very close to $b_{\text{ph}}$, which are emitted outward and inward
in the pink and purple regions of FIG. \ref{Fig:Potential-s}, respectively.
For inward-emitted photons, the Doppler effect competes with the gravitational
redshift to determine the observed frequency, leading to a blueshift
peak $\omega_{o}/\omega_{e}=1.378$ at $r_{e}=8.970$. }
\label{Fig:critical-s}
\end{figure}

Considering photons emitted with an impact parameter very close to
$b_{\text{ph}}$ in the pink and purple regions of FIG. \ref{Fig:Potential-s},
their normalized frequencies measured by observers on the celestial
sphere are plotted against the position of the star in FIG. \ref{Fig:critical-s}.
Particularly, the purple line represents photons emitted inward in
the purple region, and the pink line denotes photons emitted outward
inside the photon sphere\ in the pink region. The normalized frequency
of a photon is determined by the gravitational redshift and the Doppler
effect, which are controlled by the position and the velocity of the
photon when it is emitted, respectively. Due to light-bending near
the photon sphere, the Doppler effect can increase the observed frequency
of inward-emitted photons. At large $r_{e}$, the gravitational redshift
is weak, and hence the photons in purple line are blueshifted. The
blueshift reaches the maximum $\omega_{o}/\omega_{e}=1.378$ at $r_{e}=8.970$
and becomes $1$ at $r_{e}=3.673$. For $r_{e}<3.673$, as a result
of strong gravitational redshift near the black hole, the observed
frequency is redshifted and approaches zero as $r_{e}$ goes to $r_{h}$.

In the left panel of FIG. \ref{Fig:sky-s}, we display the normalized
frequency distribution of photons, which are emitted from the infalling
star and collected by the observers distributed on the celestial sphere.
At early times, the received photons are dominated by those emitted
in the green region of FIG. \ref{Fig:Potential-s}. Later, photons
emitted inward in the near-critical purple region start to reach the
observers after circling around the photon sphere, and come to dominate
the high-frequency observation. On the other hand, photons emitted
outward in other color regions are found to be severely redshifted
since both the Doppler effect and the gravitational redshift decrease
the observed frequency. The corresponding normalized total luminosity
are presented in the right panel of FIG. \ref{Fig:sky-s}, where a
dot corresponds to a packet of 50 photons, and the color of the dot
is that having most photons in the packet. The luminosity gradually
increases until reaching a peak at early times, and are dominated
by photons emitted in the green region roughly before $t_{o}=215$,
which is in agreement with the frequency observation. At late times,
the luminosity is mostly determined by photons with a near-critical
impact parameter emitted in the purple region. Interestingly, we find
that the luminosity decays exponentially as $e^{-\lambda t}\simeq e^{-0.140t}$,
where $\lambda$ is the Lyapunov exponent of circular null geodesics
at the photon sphere. Indeed, it was argued that the late-time decay
of the luminosity $\mathcal{L}$ is determined by the imaginary part
of quasinormal frequencies $\omega$, i.e., $\mathcal{L}\propto e^{2\operatorname{Im}\omega t}$,
in the eikonal limit \cite{Cardoso:2021sip}. For the hairy black
hole with a single-peak potential, the imaginary part of the lowest-lying
quasinormal frequencies is $-\lambda/2$ in the eikonal limit \cite{Guo:2021enm},
which gives $\mathcal{L}\propto e^{-\lambda t}$. Note that the finite
number of photons in a packet results in the non-smoothness of\ the
luminosity in FIG. \ref{Fig:sky-s}. The luminosity can become smoother
by including more photons in a packet, which, nevertheless, reduces
the accuracy of our numerical results with fixed number of total emitted
photons.

For a specific observer located at $\varphi=0$ and $\theta=\pi/2$
on the celestial sphere, the angular coordinate change $\Delta\varphi$
of light rays connecting the star with the observer is 
\begin{equation}
\Delta\varphi=2n\pi\text{,}
\end{equation}
where $n=0,1,2\cdots$ is the number of orbits that the light rays
complete around the black hole. To obtain the observational appearance
of the star seen by the observer, we focus on the collected photons
with $\cos\varphi>0.99$. We present the frequency observation in
the left panel of FIG. \ref{Fig:phio-s}, which shows a discrete spectrum
separated by the received time. The leftmost line is formed by photons
with $n=0$, which radially propagate to the observer. As expected,
the observed frequency of the $n=0$ photons decreases with the received
time because of the gravitational redshift. The rest three lines correspond
to light rays with $n=1$, $2$ and $3$ from left to right, and have
a similar shape since the emission angle is almost same for the $n\geq1$
light rays. As discussed earlier, the $n\geq1$ photons can be blueshifted
due to the Doppler effect and the strong gravitational lensing. With
a fixed $n$, the observed frequencies increase slowly, reach a maximum
and then decrease rapidly. In addition, the time delay between the
adjacent lines roughly equals to the time it takes to orbit the photon
sphere, i.e., $\Delta T\simeq2\pi b_{\text{ph}}\simeq23$\footnote{\label{ft:1} Eqn. $\left(\ref{eq:E-L}\right)$ leads to $dt/d\phi|_{r_{\text{ph}}}=b^{-1}V_{\text{eff}}^{-1}(r_{\text{ph}})=b_{\text{ph}}$,
which gives $\Delta T\simeq2\pi b_{\text{ph}}$.}. Furthermore, although the received $n=1$ photons mainly come from
the purple and pink regions of FIG. \ref{Fig:Potential-s}, those
emitted in the green and yellow regions are detected at early and
late times, respectively. Note that the angular coordinate changes
of light rays with a given impact parameter vary with the emitted
position $r_{e}$. When $r_{e}$ is small (large) enough, photons
emitted in the yellow (green) regions can circle around the black
hole once before being received. For $n\geq2$, the observer only
receive photons emitted in the purple and pink regions, which can
circle around the black hole more than once due to the near-critical
impact parameter.

The left panel of FIG. \ref{Fig:phio-s} shows the observed normalized
luminosity as a function of the observed time, which exhibits a steady
decline at early times owing to the gravitational redshift. Around
$t_{o}\simeq200$, blueshifted photons with $n=1$ start to play a
dominant role, leading to a gradual rise of luminosity until reaching
a peak. Afterwards, a sharp drop of luminosity is observed due to
strong redshifts of photons emitted at small $r_{e}$. Later, photons
with $n=2$ start to dominate the contribution to the luminosity.
The luminosity of the $n=2$ photons is about $40$ times smaller
than that of the $n=1$ photons since the $n=2$ photons are much
less than the $n=1$ photons. Despite different magnitudes, the luminosities
of the photons with $n=1$ and $2$ have a similar shape. In short,
successive receptions of photons orbiting around the photon sphere
bring about a series of observed flashes labelled by $n$. Note that
our numerical simulation does not have enough emitted photons to produce
the luminosity of the $n>2$ photons.

\subsection{Double-peak Potential}

\label{sub-sec:double peaks}

\begin{figure}[ptb]
\includegraphics[width=1\textwidth]{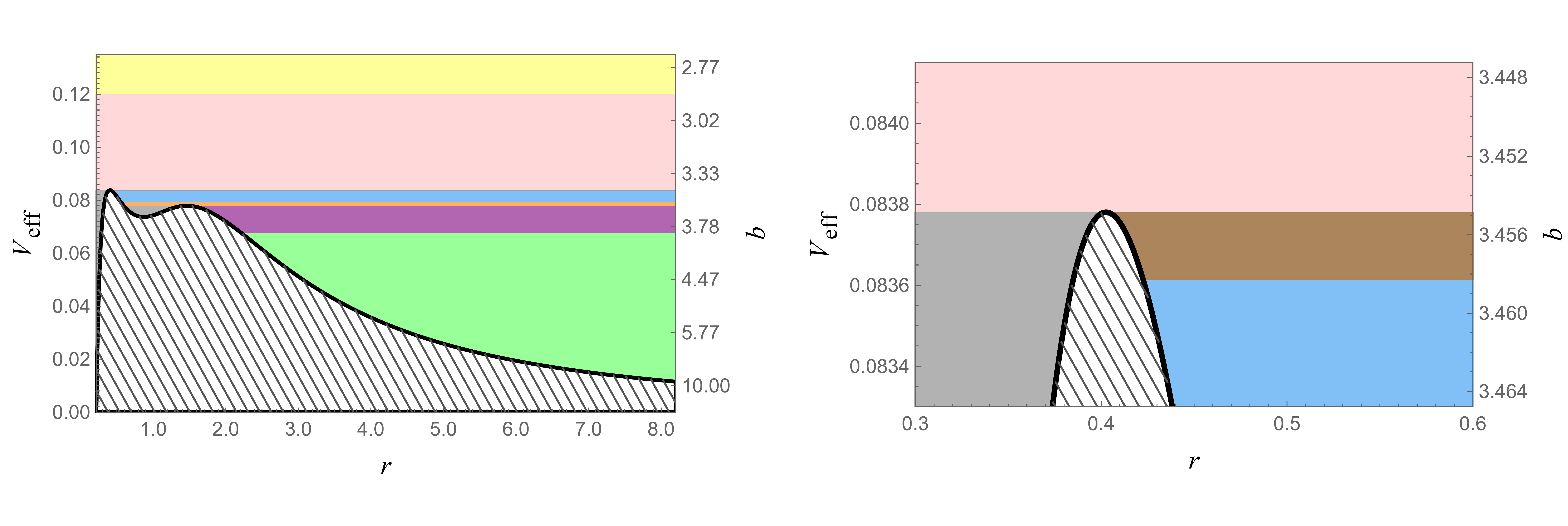}\caption{The effective potential of null geodesics in the hairy black hole
with $\alpha=0.9$ and $Q=1.064$. The potential has two peaks at
$r_{\text{in}}=0.403$ and $r_{\text{out}}=1.463$, corresponding
to the inner photon sphere with $b_{\text{in}}=3.455$ and the outer
one with $b_{\text{out}}=3.584$, respectively. The right panel highlights
the region near the inner peak. When photons are emitted outward (inward)
at $r<r_{\text{in}}$ ($r>r_{\text{out}}$) in the pink (purple) region,
they usually orbit the black hole with $\Delta\varphi\protect\geq2\pi$.
Photons emitted in the brown, blue and orange regions have $b_{\text{in}}<b<b_{\text{out}}$
and can be temporarily trapped in the region between the inner and
outer photon spheres by orbiting the black hole with $\Delta\varphi\protect\geq6\pi$.
Hence, these photons play a key role in late-time observations.}
\label{Fig:Potential-d}
\end{figure}

\begin{figure}[ptb]
\includegraphics[width=1\textwidth]{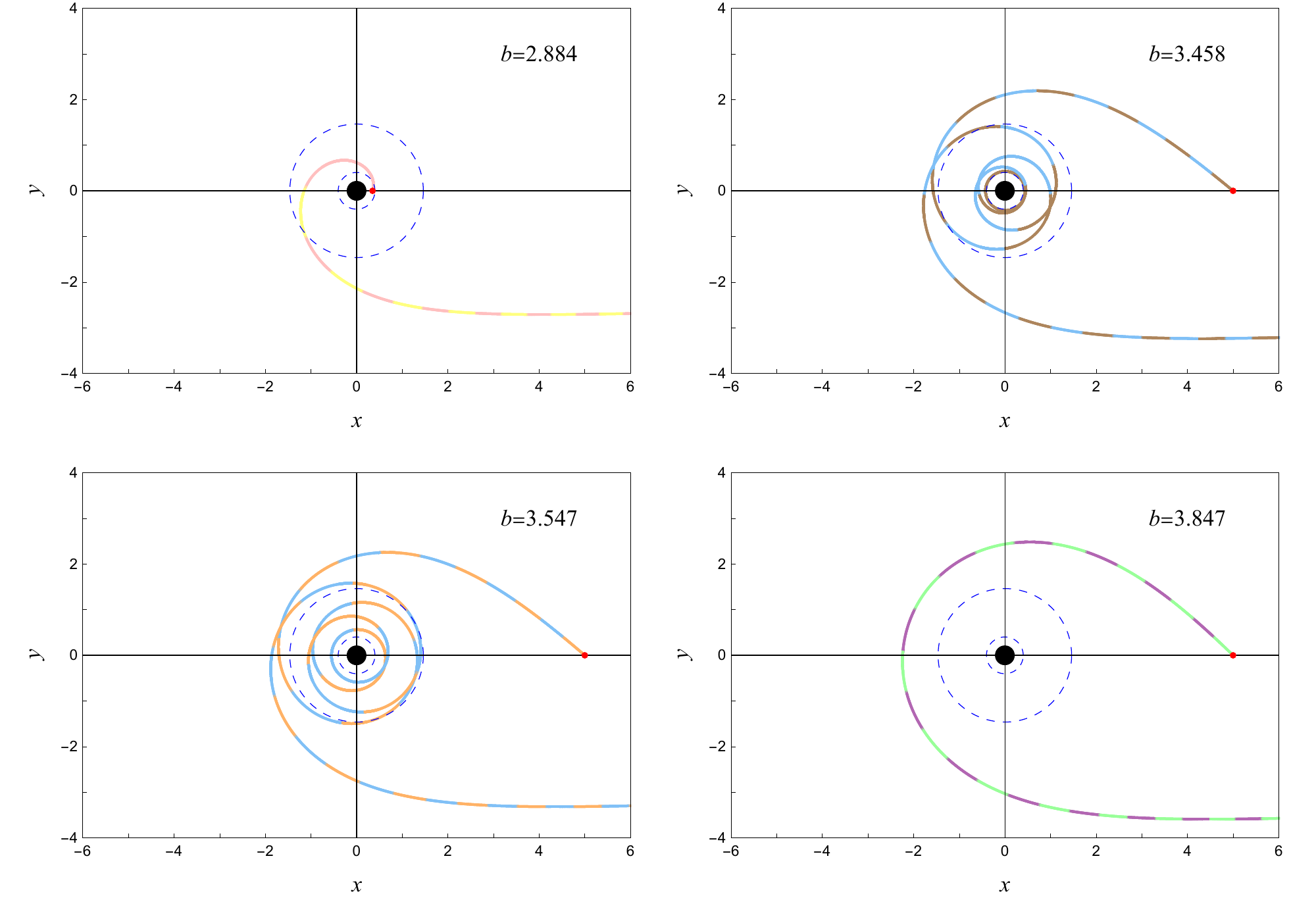}\caption{Photon trajectories in the hairy black hole with a double-peak potential.
The blue dashed circles denote the photon spheres. \textbf{Upper-Left}:
A photon is emitted at $r_{e}=0.35$ with $b=2.884$, and the light
ray has $\Delta\varphi=2\pi$. \textbf{Upper-Right}: A photon is emitted
at $r_{e}=5$ with $b=3.458$, and the light ray has $\Delta\varphi=10\pi$.
\textbf{Lower-Left}: A photon is emitted at $r_{e}=5$ with $b=3.547$,
and the light ray has $\Delta\varphi=10\pi$. \textbf{Lower-Right}:
A photon is emitted at $r_{e}=5$ with $b=3.847$, and the light ray
has $\Delta\varphi=2\pi$.}
\label{Fig:tr-d}
\end{figure}

\begin{figure}[ptb]
\includegraphics[width=1\textwidth]{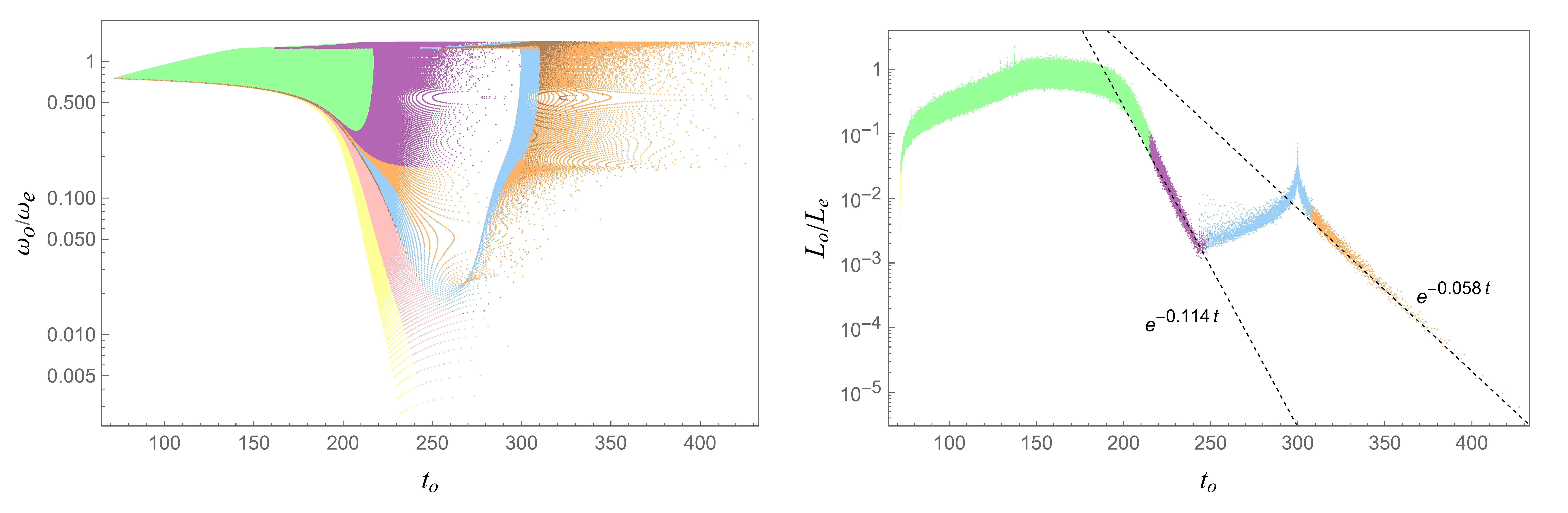}\caption{The normalized frequency distribution (\textbf{Left}) and total luminosity
(\textbf{Right}) of the infalling star measured by observers on a
celestial sphere at $r_{o}=100$ in the hairy black hole with a double-peak
potential. Before $t_{o}\simeq250$, the frequency and luminosity
observations are quite similar to the single-peak case as the outer
photon sphere plays the role of the photon sphere in the single-peak
case. For $200\lesssim t_{o}\lesssim250$, the luminosity scales as
a decay $\mathcal{L}\propto e^{-0.114t}$, in agreement with the Lyapunov
exponent $\lambda_{\text{out}}=0.11420$ at the outer photon sphere.
Interestingly, the inward-emitted photons with $b_{\text{in}}<b<b_{\text{out}}$
(i.e., those emitted in the brown, blue and orange regions) usually
linger for a long time between the inner and outer photon spheres
before being received and kick in after $t_{o}\simeq250$. For $250\lesssim t_{o}\lesssim300$,
a large number of photons with a wide range of frequencies are detected
by the observers, thereby leading to an increase of the observed luminosity
and a luminosity peak at $t_{o}\simeq300$. After $t_{o}\gtrsim300$,
less and less photons leak out the outer photon sphere, and the luminosity
decays exponentially as $\mathcal{L}\propto e^{-0.058t}$, which may
be associated with sub-long-lived quasinormal modes residing at the
outer photon sphere.}
\label{Fig:sky-d}
\end{figure}

\begin{figure}[ptb]
\includegraphics[width=1\textwidth]{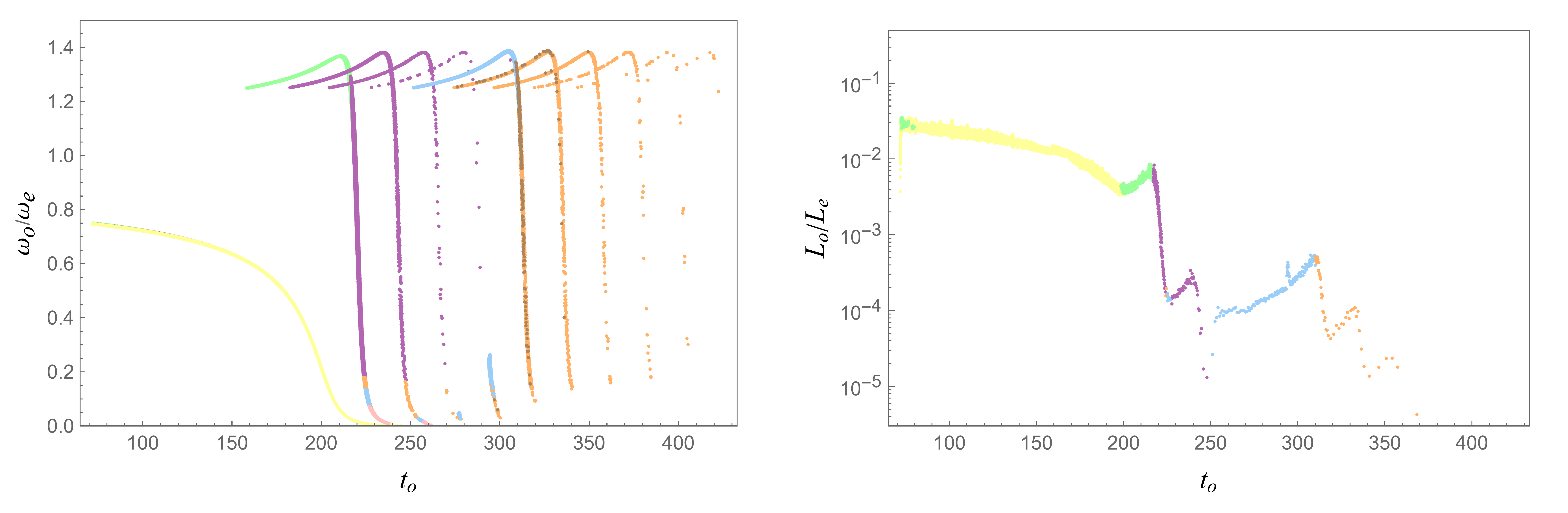}\caption{The normalized frequency and luminosity of the infalling star measured
by a far-away observer at $r_{o}=100$, $\theta=\pi/2$ and $\phi=0$
in the hairy black hole with a double-peak potential. \textbf{Left}:
The frequency lines correspond to the orbit number $n=0$, $1$, $2$,
$\cdots$ from left to right, and the $n\protect\geq1$ frequency
lines are separated by $\Delta T\simeq2\pi b_{\text{in}}\simeq2\pi b_{\text{out}}\simeq22$.
At early times, the $1\protect\leq n\protect\leq4$ frequency lines
are primarily determined by photons lingering outside the outer photon
sphere. After $t_{o}\simeq250$, photons emitted from the star successively
reach the observer after they orbit around the black hole between
the inner and outer photon spheres, causing the $n>4$ frequency lines.
\textbf{Right:} Before the photons temporarily trapped between the
photon spheres arrive, the observer sees a luminosity drop followed
by two flashes, which is quite similar to the single-peak case. After
$t_{o}\simeq250$, the received $n=5$ photons lead to a notable flash.
Subsequently, the received $n=6$ and $7$ photons give two more but
much less luminous flashes. }
\label{Fig:phi0-d}
\end{figure}

The hairy black hole with $\alpha=0.9$ and $Q=1.064$ has a double-peak
effective potential as shown in FIG. \ref{Fig:Potential-d}. The inner
peak corresponds to the inner photon sphere with the critical impact
parameter $b_{\text{in}}=3.455$ at $r_{\text{in}}=0.403$, and the
outer peak to the outer photon sphere with $b_{\text{out}}=3.584$
at $r_{\text{out}}=1.463$. Similarly, emitted photons are sorted
into seven categories by their impact parameter $b$,
\begin{itemize}
\item $b<2.884$. Yellow region in FIG. \ref{Fig:Potential-d} and yellow
dots in FIGs. \ref{Fig:sky-d} and \ref{Fig:phi0-d}.
\item $2.884\leq b<b_{\text{in}}$. Pink region in FIG. \ref{Fig:Potential-d}
and pink dots in FIGs. \ref{Fig:sky-d} and \ref{Fig:phi0-d}. In
this category, photons emitted outward inside the inner photon sphere
can circle around the inner photon sphere before reaching a distant
observer. For example, a light ray with $b=2.884$, which has $\Delta\varphi=2\pi$,
is displayed in the upper-left panel of FIG. \ref{Fig:tr-d}.
\item $b_{\text{in}}<b\leq3.458$. Brown region in FIG. \ref{Fig:Potential-d}
and brown dots in FIGs. \ref{Fig:sky-d} and \ref{Fig:phi0-d}. In
this category, photons emitted inward would circle around the inner
photon sphere roughly with $\Delta\varphi\geq10\pi$ before escaping
to the celestial sphere. For example, a light ray with $b=3.458$,
which has $\Delta\varphi=10\pi$, is displayed in the upper-right
panel of FIG. \ref{Fig:tr-d}.
\item $3.458<b\leq3.547$. Blue region in FIG. \ref{Fig:Potential-d} and
blue dots in FIGs. \ref{Fig:sky-d} and \ref{Fig:phi0-d}. In this
category, photons can be temporarily trapped between the inner and
outer photon spheres by circling around the black hole approximately
between three and five times. For example, a light ray with $b=3.547$,
which has $\Delta\varphi=10\pi$, is displayed in the lower-left panel
of FIG. \ref{Fig:tr-d}.
\item $3.547<b<b_{\text{out}}$. Orange region in FIG. \ref{Fig:Potential-d}
and orange dots in FIGs. \ref{Fig:sky-d} and \ref{Fig:phi0-d}. In
this category, if photons are emitted inward outside the outer photon
sphere, they would linger for some time around the outer photon sphere
by orbiting the black hole approximately with $\Delta\varphi\geq10\pi$.
\item $b_{\text{out}}<b\leq3.847$. Purple region in FIG. \ref{Fig:Potential-d}
and purple dots in FIGs. \ref{Fig:sky-d} and \ref{Fig:phi0-d}. In
this category, photons emitted inward outside the outer photon sphere
usually circle around the outer photon sphere more than once. For
example, a light ray with $b=3.847$, which has $\Delta\varphi=2\pi$,
is displayed in the lower-right panel of FIG. \ref{Fig:tr-d}.
\item $b>3.847$. Green region in FIG. \ref{Fig:Potential-d} and green
dots in FIGs. \ref{Fig:sky-d} and \ref{Fig:phi0-d}. 
\end{itemize}
\begin{figure}[ptb]
\includegraphics[width=0.5\textwidth]{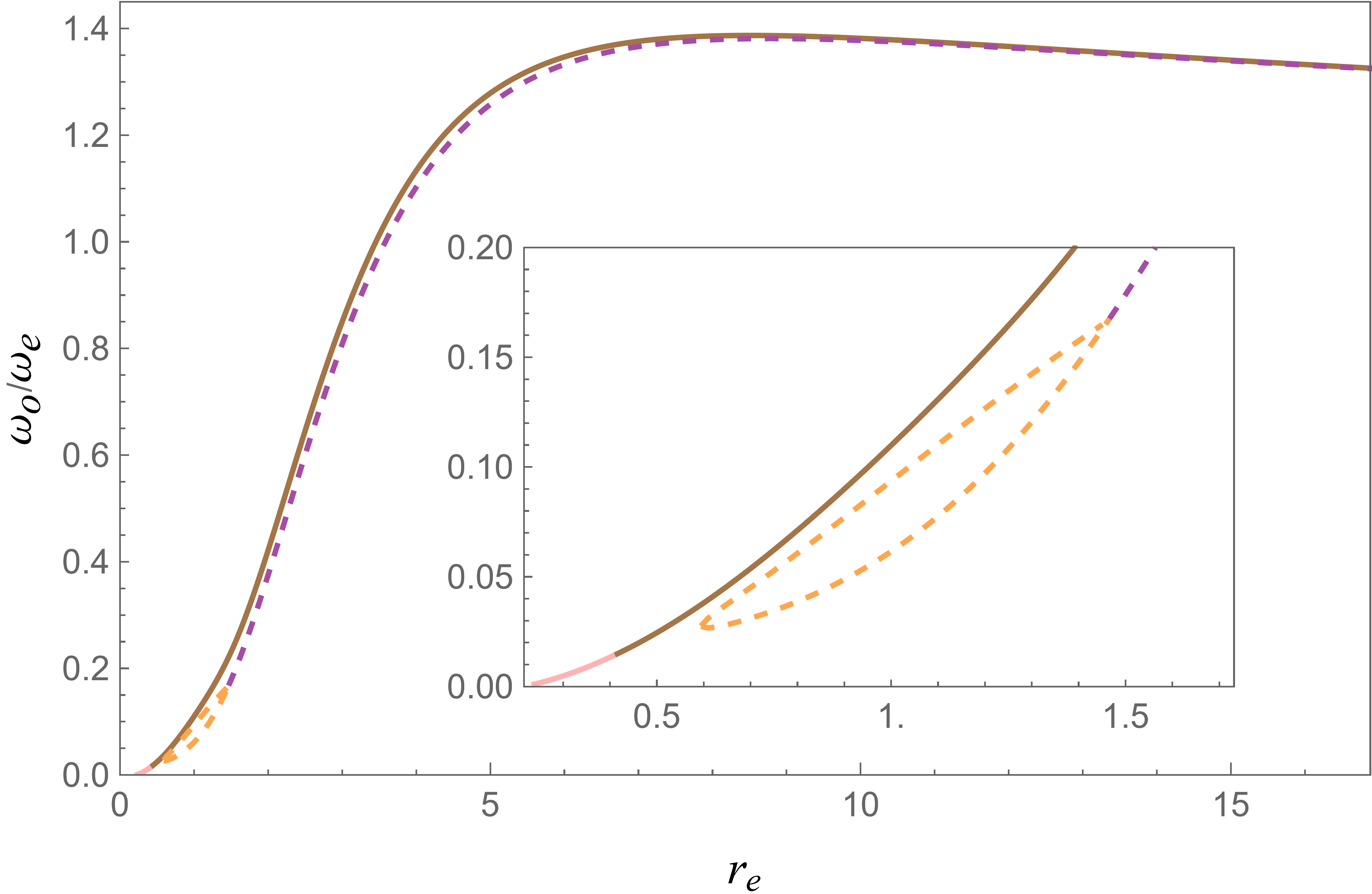}\caption{The normalized frequency $\omega_{o}/\omega_{e}$ as a function of
the emitted position $r_{e}$ for photons, whose impact parameter
$b$ is very close to $b_{\text{in}}$ (solid lines) or $b_{\text{out}}$
(dashed lines), in the hairy black hole with a double-peak potential.
For large $r_{e}$, the inward-emitted and near-critical photons are
blueshifted since the Doppler effect dominates over the gravitational
redshift. The photons emitted inward and outward between the two photon
spheres can both reach a distant observer, and denote the upper and
lower branches of the dashed line in the inset, respectively.}
\label{Fig:critical-d}
\end{figure}

In FIG. \ref{Fig:critical-d}, we plot the normalized frequency $\omega_{o}/\omega_{e}$
as a function of the emitted position $r_{e}$ for near-critical photons.
Specifically, we focus on inward-emitted photons with $b$ very close
to $b_{\text{in}}$ at $r_{e}>r_{\text{in}}$ in the brown region,
outward-emitted photons with $b$ very close to $b_{\text{in}}$ at
$r_{e}<r_{\text{in}}$ in the pink region, inward-emitted photons
with $b$ very close to $b_{\text{out}}$ at $r_{e}>r_{\text{out}}$
in the purple region and photons emitted with $b$ very close to $b_{\text{out}}$
at $r_{e}<r_{\text{out}}$ in the orange region. The colors of the
lines in FIG. \ref{Fig:critical-d} match those of the corresponding
emitted regions in FIG. \ref{Fig:Potential-d}. Moreover, photons
with $b$ very close to $b_{\text{in}}$ and $b_{\text{out}}$ are
denoted by solid and dashed lines, respectively. Similar to the single-peak
case, strong gravitational lensing around the inner and outer photon
spheres can cause blueshifts of the near-critical photons when $r_{e}$
is large enough. In particular, the normalized frequency of photons
with $b$ very close to $b_{\text{in}}$ ($b_{\text{out}}$) reaches
the maximum $\omega_{o}/\omega_{e}=1.387$ at $r_{e}=8.512$ ($r_{e}=8.784$)
and becomes the unit at $r_{e}=3.455$ ($r_{e}=3.584$). In addition,
the inset shows that the dashed line is divided into two branches
for $r_{\text{in}}<r_{e}<r_{\text{out}}$, and the upper and lower
branches correspond to inward-emitted and outward-emitted photons
emitted between the two photon spheres, respectively.

The normalized frequency distribution of photons received by the observers
distributed on the celestial sphere is presented in the left panel
of FIG. \ref{Fig:sky-d}. Roughly for $t_{o}<250$, the frequency
distribution of the received photons bears a resemblance to the single-peak
case. At early stage, most of the received photons are emitted in
the green region. Afterwards, photons that are emitted inward in the
purple region come to dominate blueshifted photons received by the
observers, and photons that are emitted outward in other colored regions
suffer from a severe redshift. Remarkably, photons emitted inward
in the blue region start reaching the celestial sphere around $t_{o}\simeq240$
after they orbit the black hole several times between the inner and
outer photon spheres. Subsequently, the observers receive photons
emitted inward in the orange and brown regions, which would linger
between the inner and outer photon spheres for a longer time. As indicated
previously, the high-frequency and low-frequency photons come from
photons emitted at large and small $r_{e}$, respectively. When $r_{e}$
is large enough, the received photons are blueshifted.

The normalized total luminosity of the infalling star is displayed
in the right panel of FIG. \ref{Fig:sky-d}, where colored dots correspond
to packets of $50$ photons. Before $t_{o}\simeq250$, the luminosity
behaves similarly to the single-peak case, i.e., a gradual rise followed
by an exponentially decaying tail as $\mathcal{L}\propto e^{-0.114t}$.
The tail is determined by photons that are emitted in the purple region
and linger outside the outer photon sphere. As expected, the Lyapunov
exponent $\lambda_{\text{out}}=0.11420$ at the outer photon sphere
controls the decay rate of the tail. It should be emphasized that
the Lyapunov exponent at the inner photon sphere is $\lambda_{\text{in}}=0.14399$,
and hence the contribution to the luminosity from photons temporarily
trapped near the inner photon sphere is suppressed. After $t_{o}\simeq250$,
photons temporarily trapped between the inner and outer photon spheres
start to play a dominant role. In particular, photons emitted in the
blue region give rise to a noticeable increase of the luminosity and
a sharp peak. Moreover, the late-time behavior is dominated by photons
emitted in the orange region and can be described by an exponential
decay, $\mathcal{L}\propto e^{-0.058t}$. In \cite{Guo:2021enm},
we found that the photons emitted in the orange region are associated
with a class of sub-long-lived quasinormal modes, thereby leading
to a more slowly decaying tail than $e^{-0.114t}$. However unlike
the single-peak case, the imaginary parts of the sub-long-lived modes
depend on the angular quantum number, and hence the decay rate of
the tail can not be universally determined.

In the left panel of FIG. \ref{Fig:phi0-d}, we exhibit the normalized
frequency of photons received by an observer located at $\varphi=0$
and $\theta=\pi/2$ on the celestial sphere. After photons are emitted
from the star, they would orbit around the black hole different times
before being captured, which results in a discrete set of the observed
frequencies indexed by the number of orbits $n$. The orbit number
$n$ of the frequency lines increases from left to right, and the
leftmost yellow line corresponds to $n=0$. As previously stated,
the $n\geq1$ frequency lines are similar and separated by the time
delay $\Delta T\simeq2\pi b_{\text{in}}\simeq2\pi b_{\text{out}}\simeq22$.
For $1\leq n\leq3$, the frequency lines are mainly contributed by
photons that are emitted in the purple region and circle around the
outer photon sphere $n$ times. When $n=4$, photons emitted in the
purple region are rarely seen, and some observed low-frequency photons
are from emission between the photon spheres in the blue and orange
regions. Inward-emitted photons in the blue, orange and brown regions
can circle around the black hole $5$ times between the inner and
outer photon spheres and come back to start arriving after $t_{o}\simeq70\times2+22\times5\simeq250$,
which forms the $n=5$ frequency line. Later, photons orbiting around
the black hole more than $5$ times successively reach the observer
and give rise to the $n>5$ frequency lines.

The normalized luminosity of the star measured by the observer is
displayed in the right panel of FIG. \ref{Fig:phi0-d}. Before $t_{o}\simeq250$,
the luminosity observation resembles that in the single-peak case,
i.e., two flashes following the decrease of the luminosity. Owing
to the leaking of photons temporarily trapped between the photon spheres,
a noticeable flash appears around $t_{o}\simeq300$, and two much
fainter flashes are seen later on. These three flashes are caused
by the received photons orbiting around the black hole $5$, $6$
and $7$ times, respectively. For $n>7$, we do not have sufficient
photons to compute the luminosity.

\section{Conclusions}

\label{sec:CONCLUSIONS}

In this paper, we investigated observational appearances of a point-like
freely-falling star, which emits photons isotropically in its rest
frame, in hairy black holes, which possess one or two photon spheres
outside the event horizon. In particular, we considered the frequency
and luminosity of received photons measured by all observers on a
celestial sphere and a specific observer. Interestingly, it was found
that the existence of an extra photon sphere can significantly change
the late-time observations, which may provide a new tool to detect
black holes with multiple photon spheres.

For hairy black holes with a single-peak potential, emitted photons
can be temporarily trapped just outside the photon sphere and subsequently
reemitted, resulting in several interesting observational signatures.
First, photons with a near-critical impact parameter can be blueshifted
since the photon sphere acts as a reflecting wall (see FIGs. \ref{Fig:critical-s}
and \ref{Fig:sky-s}). Second, at late times, the total luminosity
measured by observers on the celestial sphere decreases exponentially
with time, which is controlled by the Lyapunov exponent of circular
null geodesics at the photon sphere (see FIG. \ref{Fig:sky-s}). Last
but not least, photons orbiting around the photon sphere different
times produce a cascade of flashes observed by the specific observer.
The luminosity of the flashes decreases sharply with the orbit number,
and the frequency contents of the flashes are similar (see FIG. \ref{Fig:phio-s}).
It is worth emphasizing that the observational appearances in the
hairy black holes with a single-peak potential are similar to those
in Schwarzschild black holes reported in \cite{Cardoso:2021sip}.

When an extra photon sphere appears outside the event horizon, we
found that light rays can get trapped between the two photon spheres
longer than just outside the photon spheres, which appreciably impacts
late-time observations. In fact, due to arrivals of photons trapped
between the photon spheres, the total luminosity rises to a peak,
which is succeeded by a slow exponential decay (see FIG. \ref{Fig:sky-d}).
This slow decay may be related to sub-long-lived quasinormal modes
living near the outer photon sphere, as found in \cite{Guo:2021enm}.
The specific observer would see two cascades of flashes, which contain
a similar frequency content. While the earlier cascade is mainly determined
by photons orbiting outside the outer photon sphere, photons trapped
between the photon spheres primarily give rise to the later one (see
FIG. \ref{Fig:phi0-d}). In summary, black holes with multiple photon
spheres will lead to distinctive optical appearances of an infalling
star. In the future studies, it will be of great interest if our analysis
can be generalized to more astrophysically realistic models.
\begin{acknowledgments}
We are grateful to Qingyu Gan and Xin Jiang for useful discussions
and valuable comments. This work is supported in part by NSFC (Grant
No. 12105191, 11947225 and 11875196). Houwen Wu is supported by the
International Visiting Program for Excellent Young Scholars of Sichuan
University. 
\end{acknowledgments}

 \bibliographystyle{unsrturl}
\bibliography{ref}

\end{document}